\documentclass[12pt]{JHEP}
\usepackage{amsmath, amsfonts,euscript,bbm}
\usepackage{epsfig, cite}
\usepackage{ifthen}
\usepackage{graphicx}
\usepackage{psfrag}

\newboolean{Notes}\setboolean{Notes}{true}
\newcommand{\notes}[1]
	    {\ifthenelse{\boolean{Notes}}{{\tt #1}}{}}

\newcommand{\bas}{\begin{eqnarray*}}
\newcommand{\eas}{\end{eqnarray*}}
\newcommand{\ba}{\begin{eqnarray}}
\newcommand{\ea}{\end{eqnarray}}
\newcommand{\dd}{\Delta}
\newcommand{\oo}{\Omega}
\newcommand{\e}{\epsilon}

\def\h{h_A^{2}} 

\def\F{\tilde F} 
\def\t{\theta}
\def\w{\wedge}
\def\l{\lambda}
\def\hf{\frac12}
\def\ap{{\alpha^{\prime}}}

\title{The (p,q) String Tension in a Warped Deformed Conifold}
\author{Hassan Firouzjahi \footnote{firouzh@lepp.cornell.edu} , Louis Leblond \footnote{lleblond@lepp.cornell.edu} \,
and S.-H. Henry Tye \footnote{tye@lepp.cornell.edu}\\
  Newman Laboratory for Elementary Particle Physics, Cornell University\\
   Ithaca, NY, 14853}

\abstract{We find the tension spectrum of the bound states of $p$ fundamental strings and $q$ D-strings at the bottom of a warped deformed conifold. We show that it can be obtained from a D3-brane wrapping a 2-cycle that is stabilized by both electric and magnetic fluxes.  
Because the F-strings are $Z_{M}$-charged with non-zero binding energy, binding can take place even if $(p,q)$ are not  coprime. Implications for cosmic strings are briefly discussed. 
} 


\preprint{}

\begin{document}

\section{Introduction}

Fundamental string (F-string) tension in 10 dimensions defines the string scale $\ap$ via  $T_{F1}=1/2 \pi \ap$. In type IIB theory, there are branes including D1-branes, or D-strings, with tension
$T_{D1}=1/2 \pi \ap g_{s}$, where $g_{s}$ is the string coupling.  In light of all the progress coming from dualities in string theory, we now know that the D-strings and the F-strings should be considered on the same footing and a general string state in type IIB is the bound state of these two types of strings.
In 10 flat dimensions, supersymmetry dictates that the tension of the bound state of $p$ F-strings and $q$ D-strings is given by \cite{Schwarz:1995dk},
	\ba\label{flat}
	T_{{p,q}} = T_{F1} \sqrt{p^2  +\frac{q^2}{g_s^2}}\, .
	\label{pqtension10}
	\ea
However, if our universe is described by superstring theory, six of the spatial dimensions must be compactified. In type IIB string theory, our universe can be described as a brane world scenario with flux compactification \cite{Giddings:2001yu,Kachru:2003aw}. 
In such a scenario, the standard model particles are likely to be light open string modes in a warped throat present in the Calabi-Yau manifold. It will be interesting to know how the above tension formula (\ref{pqtension10}) is modified in such a throat.

The string tension formula  at the bottom of a throat is also interesting from another perspective. Brane inflation, a particularly attractive scenario of the inflationary universe, involves the D3-anti-D3-brane annihilation towards the end of inflation \cite{Dvali:1998pa,Kachru:2003sx}. 
This annihilation releases the energy to heat the universe to start the hot big bang era. Presumably, anything that is not forbidden will be produced during this annihilation, in particular F-
and D-strings. These strings with cosmological sizes will appear as cosmic strings \cite{Jones:2002cv, Sarangi:2002yt, Copeland:2003bj, Becker:2005pv}.
Their range of tension is compatible with the current observational bounds \cite{Firouzjahi:2005dh,Shandera:2006ax}. 
At the same time, their values are close enough to be detected in the near future, for example, via gravitational wave detectors, 
and the study of their properties may be the long sought experimental window into string theory. Since the D3-anti-D3-brane annihilation most likely takes place at the bottom of a throat, that will be where the cosmic superstrings are. The $(p,q)$ cosmic string tension spectrum in the throat will have interesting phenomenological and cosmological implications.
This is a very strong motivation to understand cosmic superstrings better and our goal in this note is to determine the tension spectrum of the $(p,q)$ string at the bottom of a throat in type IIB string theory.  

To be specific, we consider the Klebanov-Strassler (KS) throat \cite{Klebanov:2000hb}  whose properties are relatively well understood. On the gravity side, this is a warped deformed conifold. Inside the throat, the geometry is a shrinking $S^{2}$ fibered over an $S^{3}$.
The tensions of the bound state of $p$ F-strings and that of $q$ D-strings were individually computed for the KS throat  \cite{Herzog:2001fq, Hartnoll:2004yr, Gubser:2004qj} to be
\begin{align}
\label{pqseparate}
T_{F1}& \simeq \frac{\h}{2 \pi \ap} \frac{b M}{\pi} \sin \left(\frac{\pi p}{M}\right), & T_{D1} &=  \frac{\h}{2 \pi \ap}\frac{q}{g_s},
\end{align}
where $p,q$ are integers, $h_A$ is the warp factor at the bottom of the throat, $b=0.93$ is a number numerically close to one and  $M$ is the number of fractional D3-branes, that is, the units of 3-form RR flux $F_3$ through the $S^{3}$. (This $T_{F1}$ formula yields values that are within a percent of that
obtained from the more precise formula in Ref\cite{Herzog:2001fq}.)
Very interestingly, the fundamental strings are charged in $\mathbb{Z}_M$ and are non-BPS. 
The D-string on the other hand is charged in $\mathbb{Z}$ and is BPS with respect to each other.  

The interpretation of these strings in the gauge theory dual is known. The fundamental string is dual to a confining string between a quark and an anti-quark \cite{Douglas:1995nw,Hanany:1997hr,Callan:1999zf}. 
Since it is a convex function, i.e., $T_{p+p'} < T_{p} + T_{p'}$, the $p$-string will not decay into strings with smaller $p$. Since $M$ quarks can form a baryon, the bound state of $M$ number of F-strings has zero tension; that is, $M$ of them can terminate at a point, i.e., the baryon. On the other hand, the D-string is dual to an axionic string.  Indeed, it is argued that the gauge theory dual of the KS throat must contain a pseudo-scalar bound state (glueball) that plays the role of an axion field \cite{Gubser:2004qj}. This axion field is the Goldstone boson associated to the spontaneous breaking of a global $U(1)$ baryon symmetry.  Upon compactification, the global symmetry becomes local and the string defects coming from the spontaneous breaking of the symmetry are the Abelian Higgs vortices. 
It seems the D-strings are BPS with respect to each other \cite{Gubser:2004qj}. 

On the gravity side, these tensions were calculated by considering a D3-brane stabilized on a 2-cycle by  magnetic flux \cite{Bachas:2000ik}. This is closely related to the dielectric effect \cite{Myers:1999ps}. Although both the 5-form flux $F_{5}$ and the NS-NS 2-form field $B_{2}$ vanish at the bottom of the throat, the $F_{3}$ flux blows up the $p$ F-strings into a D3-brane wrapping a finite $S^2$ inside the $S^3$. In Ref.\cite{Herzog:2001fq}, the tension is obtained by considering the F-string as the S-dual of a blow-up D1-brane where the D3-brane with $p$ units of magnetic flux is wrapping a finite $S^2$ inside the $S^3$. On the other hand,  
it was shown that $q$ D-strings can be seen as a D3-brane with $q$ units of magnetic flux wrapping a shrinking $S^2$ cycle. Ref.\cite{Gubser:2004qj} finds that the D-strings are BPS, that is, no sign that the D-strings are $\mathbb{Z}_K$ valued, where $K$ is the value of the NS-NS 3-form flux.

Note that the F-string tension and the D-string tension are evaluated using two different approaches.
To find the $(p,q)$ string tension, we like to start with a unified picture that allows the evaluation of either the F-string or the D-string tension. This is easy to achieve. The F-string can be seen as a D3-brane with electric flux. We shall show that the electric flux stabilizes a D3-brane the same way the magnetic flux does. So the S-dual step is bypassed. This unified description allows us to very simply describe the general $(p,q)$ bound states by turning on simultaneously both electric and magnetic fluxes on a D3-brane.  A discussion of confining strings as electric flux tubes in a different context can be found in \cite{Herzog:2002ss}. One can also see F-string as electric flux tubes in the D3-anti-D3 action \cite{Bergman:2000xf, Cho:2005wp}.

All of this is expected from the $SL(2,\mathbb{Z})$ duality of Type IIB theory. 
Following from the $SL(2,\mathbb{Z})$ duality transformation that mixes the $B_2$ and $C_2$ fields, a F-string is S-dual to a D-string. 
The D3-brane is self-dual under $SL(2,\mathbb{Z})$ duality \cite{Tseytlin:1996it}. On the other hand, 
the electric (E) and magnetic (B) fluxes on the D3-brane are mixed by the $SL(2,\mathbb{Z})$ duality transformation \cite{Gibbons:1995cv,Gibbons:1995ap,Green:1996qg}.  
To obtain the $p$ F-strings tension, Ref.\cite{Herzog:2001fq} starts with a D3-brane with $p$ units of magnetic flux to obtain $p$ D-strings and then S-dual it to $p$ F-strings.
Alternatively, one can S-dualize the D3-brane so it becomes a D3-brane with electric fluxes to obtain $p$ F-strings directly. Or even simpler, we can simply start with a D3-brane with the appropriate electric fluxes directly. 


In this note, we just start with appropriate electric and magnetic fluxes on a D3-brane. This is analogous to the early attempts of getting the $(p,q)$ string tension by dissolving electrical fluxes on a D-string \cite{Witten:1995im,Li:1995pq}. These methods could only yield approximately the $(p,q)$ string tension because the gauge dynamics on $n$ D-strings is non-abelian and the gauge coupling is relevant in two dimensions leading to a strongly coupled gauge theories \cite{Witten:1995im}. Fortunately, our method completely avoids this problem since the gauge theory on the D3-brane is always $U(1)$ with a dimensionless coupling constant. In some sense, we get rid of the non-abelian physics by doing the blow-up.

In this approach, we find that the tension formula for the bound states turns out to have a simple (expected) form:
\ba\label{finalanswer}
T_{p,q} \simeq  \frac{\h}{2 \pi \ap} \sqrt{  \frac{q^2}{g_s^2} + 
(\frac{b M}{\pi})^2 \sin^2(\frac{\pi p}{M})},
\ea  
but the way it comes about is interesting.  
Indeed, the tension is obtained by minimizing the Hamiltonian of the D3-brane world volume action after integrating out the extra dimensions. Care must be taken with the Hamiltonian when one has an electric field on a D-brane.  For example the Chern-Simons terms which do not contribute to the stress energy tensor due to their topological nature nevertheless affect the Hamiltonian (hence the tension and energy) by coming into play via the conjugate momentum.  This contribution turns out to be crucial here and leads to the above simple formula (\ref{finalanswer}).  

This formula has the right limits.
Setting either $p=0$ or $q=0$ reproduces Eq.(\ref{pqseparate}).
For $M \rightarrow \infty$ and $b=h_{A}=1$, it reduces to Eq.(1.1). Because $p$ is $Z_{M}$-valued with non-zero binding energy, binding can take place even if $(p,q)$ are not  coprime. Also, $M$ fundamental strings can terminate to a point-like baryon, irrespective of the number of D-strings around. 

The paper is divided as follows. Sec. 2 reviews the properties of the KS throat we need. Sec. 3 contains the calculation and the main result. Sec. 4 includes general discussions and some comments on the
issues remaining.

\section{A Throat in the Calabi-Yau Manifold}

\subsection{The Conifold}
A cone is defined by the following equation in ${\cal{\bf{C^4}}}$
\ba
\label{6cone}
\sum_{i=1}^{4} w_i^2=0
\ea
Here Eq.(\ref{6cone}) describes a smooth surface apart from the point $w_i=0$. 
The geometry around the conifold is studied in \cite{Candelas:1989js}.
The base of the cone is a manifold $X$
given by the intersection of the space of solutions of Eq.(\ref{6cone}) with a sphere of radius $r$ in ${\cal{\bf{C^4}}}=R^8$, $$\sum_i |w_i|^2 = r^2$$

We are interested in Ricci-flat metrics on the cone which in turn imply that the base of the conifold is a 
Sasaki-Einstein manifold.  The simplest five dimensional Sasaki-Einstein manifold for $N=1$ supersymmetry is $T^{1,1}$ and it is the only manifold for which the deformation is explicitly known \cite{Klebanov:2000hb}.

The metric on the conifold with base $T^{1,1}$ is 
\ba
\label{6t11}
ds_{6}^{2}&=& dr^{2} + r^{2} ds_{T^{1,1}}^2 \,   \\
ds_{T^{1,1}}^2 &=&\frac{1}{9}(\,d\psi +\sum_{i=1}^{2} 
\cos \theta_i\, d \phi_i\, )^2 + \frac{1}{6}\sum_{i=1}^{2}
 (\, d\theta_i^2 +  \sin^2 \theta_i\, d\phi_i^2\,)\, . \nonumber  
\ea
It can be shown that $$T^{1,1}=(SU(2)\times SU(2))/U(1)=S^3 \times S^3/U(1)$$ 
which has topology of $S^2 \times S^3$ (with $S^2$ fibered over $S^3$). 
If $\varphi_1$ and $\varphi_2$ are the two Euler angles of the two $S^{3}$s, 
respectively, then their difference corresponds to $U(1)$ while $\psi=  \varphi_1+ \varphi_2$. 
Since $2 \pi \ge \varphi_i \ge 0$, the range of $\psi$ is $[0, 4 \pi]$.

\subsection{The Warped Deformed Conifold}

The Klebanov-Strassler throat that we are interested in is actually a warped deformed conifold. This warped deformed conifold emerges in the presence of fluxes.
The R-R flux $F_3$ wraps the $S^{3}$ while NS-NS flux $H_3$ wraps the dual 3-cycle $B$ that generates the warped throat, with warp factor $h(r)$. 
\ba
\label{KMNhA}
\frac{1}{ 4\pi^{2}\ap}\int_B H_3 &=& -K, \quad \quad 
\frac{1}{ 4\pi^{2} \ap}\int_{S^{3}} F_3 = M 
\ea
Geometrically, the conical singularity of Eq.(\ref{6cone}) can be removed by replacing the 
apex by an $S^3$ \cite{Candelas:1989js},
\ba
\sum_{i=1}^{4} w_i^2=\epsilon^2
\ea
where we shall take $\epsilon$ to be real and small.
It will be convenient to work in a diagonal basis of the metric, given by
the following basis of 1-forms \cite{Minasian:1999tt,Herzog:2001xk},
\ba
g^1\equiv\frac{e^1-e^3}{\sqrt{2}}~~~~~,~~~~~
g^2\equiv\frac{e^2-e^4}{\sqrt{2}}\nonumber\\
g^3=\frac{e^1+e^3}{\sqrt{2}}~~~~~,~~~~~
g^4\equiv\frac{e^2+e^4}{\sqrt{2}}\nonumber\\
g^5\equiv e^5
\ea
where
\ba
\label{angular}
e^1&\equiv& -\sin \theta_1\, d\phi_1 ~~~~~,~~~~~ e^2\equiv d\theta_1 \, ,
 \nonumber\\
e^3&\equiv& \cos \psi \sin \theta_2\, d \phi_2 -\sin \psi\, d \theta_2\, , 
\nonumber\\
e^4&\equiv& \sin \psi \sin \theta_2\, d \phi_2 +\cos \psi\, d \theta_2\, ,
\nonumber\\
e^5&\equiv& d \psi +\cos \theta_1\, d\phi_1 + \cos \theta_2\, d \phi_2
\ea
The metric of the deformed conifold is studied in 
\cite{Minasian:1999tt,Ohta:1999we,Herzog:2001xk}
\ba
\label{deformedmetric}
ds_6 ^2 = \frac{1}{2}\e^{4/3} K(\tau)\left[\frac{1}{3K^3(\tau)}
(d\tau^2+ (g^5)^2)+\cosh^2(\frac{\tau}{2})[(g^3)^2+(g^4)^2]\right.\nonumber\\
+\left. \sinh^2(\frac{\tau}{2})[(g^1)^2+(g^2)^2]\right]
\ea
where
\ba
\label{K(tau)}
K(\tau)=\frac{(\, \sinh(2\tau)-2\tau\,)^{1/3}}{2^{1/3}\sinh \tau} \, 
\ea
At $\tau \to 0$, the $S^{2}$ (the $g^1$ and $g^2$ components of the metric) shrinks to zero,
and we are left with 
\ba
d\oo_3^2=\frac{1}{2}\e^{4/3}(2/3)^{1/3}\, \left(\frac{1}{2}(g^5)^2
+(g^3)^2+(g^4)^2 \right)
\ea
which is a spherical $S^3$.
Turning on fluxes, the ten-dimensional metric takes the following ``warped'' form
\ba
\label{10Dwarp}
ds_{10}^2 = h^{-1/2}(\tau) \eta_{\mu \nu}\,dx^{\mu} dx^{\nu}+
h^{1/2}(\tau)\, ds_6^2
\ea
where $ds_6^2$ is given above in Eq.(\ref{deformedmetric}). 
The warp factor $h(\tau)$ is given by the following integral expression \cite{Klebanov:2000hb}
\ba
\label{h(tau)}
h(\tau)= 2^{2/3}\,(g_s M \alpha')^2\, \e^{-8/3}\, I(\tau)\, ,
\ea
\ba
\label{I(tau)}
I(\tau)\equiv\int_{\tau}^{\infty} d\, x \frac{x\coth\, x -1}{\sinh^2 \,x }
\left(\,\sinh(2x)-2x\, \right)^{1/3} \, .
\ea
We use the convention that the warp factor $h_{A}$ at the bottom of the throat is
\ba
h_{A}= h(0)^{-1/4}=\e^{2/3}2^{-1/6} a_0^{-1/4} (g_s M \alpha')^{-1/2}
\ea
where $a_0 \equiv I(\tau=0) \sim 0.71805$. So $h_{A}$ measures the mass scale at the bottom of the throat relative to that in the bulk. In a compact manifold it can be related to the parameters K, M and $g_s$ by $ h_{A}=e^{-2\pi K/3g_{s}M}$.

\section{The Spectrum of $(p,q)$ Strings from a Wrapped D3-brane}
\label{sectiongen}

We are interested in string-like objects extending in the $x^0$ and $x^1$ directions in the 4 dimensional spacetime coming from a wrapped D3-brane. We first determine the flux configuration that gives $p$ units of F-string charge and $q$ units of D-string charge and then we proceed to find their tensions.

\subsection{Charges of a $(p,q)$ string}

We are looking for a flux configuration on a D3-brane that gives $p$ units of F-string charge and $q$ units of D-string charge. To achieve this we wrap our D3-brane on a 2-cycle in the extra dimensions. 
We choose the following gauge for the world volume coordinates $\xi^0 = x^0$, $\xi^1 = x^1$ while $\xi^{2}$ and $\xi^{3}$ are the coordinates for the 2-cycle.

The equation of motion for $C_2$ and $B_2$ in type IIB with a D3-brane source are
\begin{align*}
d\star \tilde F_3 & = F_5\w H_3 + \frac{\delta S_{D_3}}{\delta C_2}\wedge \delta^{6}({\bf x}),\\
d\star (e^{-2\phi} H_3 -  C_0 \tilde F_3) & =  F_5\w F_3 + \frac{\delta S_{D_3}}{\delta B_2}\wedge \delta^{6}({\bf x}),
\end{align*}
where $\tilde F_3 = F_3-C_0H_3$.  One can obtain the charge by integrating the right hand side of these equations over an $S^8$ surrounding the $x^0, x^1$ directions.
\begin{align}\label{charge}
Q_{D1} & = \int_{S^8} d\star \tilde F_3 & Q_{F1} & =  \int_{S^8} d \star (e^{-2\phi} H_3 - C_0 \tilde F_3)\nonumber\\
& = \int_{S^8}F_5\w H_3 + \int_{S^2} \frac{\delta S_{D_3}}{\delta C_2} 
& &= \int_{S^8}F_5\w F_3 + \int_{S^2} \frac{\delta S_{D_3}}{\delta B_2}
\end{align}
For both cases, the first term measures the D1/F charge of the background itself.  Here we are only interested in the charges induced by the D3-brane and so we will drop them.  From now on, we consider solutions for which the dilaton is constant $e^\phi = g_s$.

To calculate the D3-brane induced charges, the D3-brane action is needed
\ba
\label{action}
S_{D3}= -\frac{\mu_3}{g_s}\, \int d^4 \xi\sqrt{-| g_{a b} +{\cal{F}}_{ab} |} + 
\mu_3 \int_{\mathbb{R}^2\times {\cal{M}}^2} \left(C_2 \wedge {\cal{F}} + \hf C_0{\cal{F}}\wedge{\cal{F}}\right)\, ,
\ea
where ${\cal{F}}_{ab}= B_{ab} + \lambda F_{ab}$, $\lambda= 2\pi\, \alpha'$ and
$\mu_3= 1/{(2 \pi)^3\, \ap^2}$ is the D3-brane charge.  We have set all the scalars (except $C_0$) to zero.  Note that the induced D-string charge from the D3-brane
\ba
\mu_3 \int_{S^2} {\cal{F}} \nonumber
\ea
 is not quantized \cite{Bachas:2000ik}.  Only the sum of the induced charge from the D3-brane and the induced charge from the background is quantized \cite{Taylor:2000za}. The way to understand this is to remember that $\cal{F}$ contains the pull-back of $B_2$ that could in principle vary continuously.  Only after solving consistently the type IIB equations of motion with a D3-brane source would we find that the two terms in (\ref{charge}) add up to quantized integers.

Nevertheless it was shown that $F_{23}$ is quantized \cite{Bachas:2000ik}. Therefore, the induced D-string charge from the D3-brane is quantized if the pull-back of the NS-NS 2 form is 
either quantized or vanishes on $S^2$.  The latter is the case at the tip of the original KS background 
but not in its S dual.
So, in the case where the pull-back of $B_2$ on $S^2$ is zero, we have $q$ units of D-string charge when $F_{23} = \frac{q}{2} \sin \theta\, d\theta d\phi$ using that $2\pi \l \mu_3 =\mu_1 = 1/\l$. 
And if the pull-back is non-zero then the $F_{23}$ flux is still the same but the D-string charge (coming from the D3-brane)  is no longer quantized.  

For the F-string, we can relate the charge to the conjugate field strength.  Indeed, since the action only depends on the gauge invariant quantity ${\cal{F}}$, we have that 
\begin{align*}
\frac{\delta S_{D3}}{\delta B_2} & = \frac{1}{\l}\frac{\delta S_{D3}}{\delta F}
\end{align*}
The conjugate field (i.e., the displacement field, or the dual field strength in the Montonen-Olive sense) of the electric field $E_i=F_{i0}$ is given by
\ba
\F^{\mu\nu}  = - \frac{\delta S}{\delta F_{\mu\nu}}
\ea
Because of the complicated form of the DBI action, it is simpler to write $\F$ as a scalar (not a 2 form).  The D3-brane induced F-string charge is
\ba
-\frac{1}{\l} \int \F^{01}\sin\t d\t d\phi.
\ea
There is no subtlety with quantization and one gets $p$ units of F-string charge 
 when  $\F^{01} = -\frac{p}{4\pi}$. It is important to remember that $\F^{\mu\nu}$ is from the point of view of the gauge theory as fundamental as $F^{\mu\nu}$ and like the latter it is quantized.

In summary, the flux configuration on a D3 wrapping a 2-cycle that can induce the charge of $p$ F-strings and $q$ D-strings is
\begin{align}
\F^{01} & = -\frac{p}{4\pi} & F_{23} & = \frac{q}{2} 
\end{align}

\subsection{The Hamiltonian of a D3-brane}

Now that we know which flux configurations give the correct  charge
of $p$ F-strings and $q$ D-strings individually, we would like to calculate the tension
of bound state of $(p,q)$ strings.
For this purpose one needs to calculate the Hamiltonian of the system of D3-brane with magnetic 
and electric fluxes turned on in its world volume.

We start with a metric of the following form 
\ba
ds^2 = h^{2}\eta_{\mu \nu} dx^{\mu} dx^{\nu} + ds_6^2 \, ,
\ea
where $h$ is the warp factor which is a function of the internal coordinates and it should not be confused with the convention used in Eq.(\ref{10Dwarp}). Our D3-brane still spans the $x^0$, $x^1$ direction and a 2-cycle in the extra dimensions. At this point we will make some assumptions. Let us suppose that only $B_{23} \neq 0$ as this is very often the case. We assume that the magnetic and electric field are parallel to each other so we turn on only $F_{23}$ and $F_{01}$. We will also assume that $F_{01}$ does not depend on the extra dimensions coordinates.
Then we have 
\ba
g_{ab} + {\cal{F}}_{ab} & = \left( 
\begin{array} {cccc} 
-h^{2} & -\l F_{10} & 0 & 0\\
\l F_{10} & h^{2} & 0 & 0\\
0& 0 & g_{22} & {\cal F}_{23}\\
0 & 0 & -{\cal F}_{23} & g_{33}
\end{array}\right)
\ea
The D3-brane action (\ref{action}) becomes
\ba
\label{def}
S =  \int dx^0dx^1 \left[\, -\dd \sqrt{h^{4} - \l^2F_{10}^2} - \oo F_{10} \right] \, ,
\ea
where
\ba
\dd  \equiv \frac{\mu_3}{g_s} \int d^2\xi \, (g_{22}\, g_{33} +  {{\cal F}_{23}}^2)^{\hf} \, ,\nonumber\\
\oo  \equiv \l \mu_3 \int d^2\xi\left[ \, (C_2)_{23} + C_0 {\cal{F}}_{23} \right]\, .\nonumber
\ea

To get the Hamiltonian one needs to calculate the conjugate momentum D (here D has one upper index which we drop).
\ba
\label{D}
D  = - \frac{\delta {\cal{L}}}{\delta{F_{10}}} =  \frac{-\dd \lambda^2 F_{10}}{\sqrt{h^{4} -\lambda^2 F_{10}^2}} + \oo
\ea 
Note that $D$ is related to $\F$ defined in the previous subsection through a factor equal to the volume of the $S^2$ sphere, $D = 4\pi \tilde F^{10} = p$.  The factor of $4\pi$ is because the lagrangian we are using to calculate $D$ is obtained after we integrate out the 2-cycle (it is a 2--dimensional lagrangian). 
One can then solve for the electric flux in terms of D, 
\ba
\label{F10}
F_{10} = - \frac{h^{2}(\frac{D-\oo}{\lambda^2 \dd}) }{\sqrt{1+ (\frac{D-\oo}{\lambda \dd})^2}}\, . 
\ea
The two dimensional Hamiltonian therefore is
\ba
\label{Ham}
{\cal{H}} &=& D F_{01} - {\cal{L}} \nonumber\\
& =& \frac{h^{2}}{\l}\sqrt{\Delta^2\l^2 + (D-\oo)^2}
\ea
The value of this Hamiltonian after minimization is the $(p,q)$ string tension.

\subsection{The KS Throat}

We will focus at the tip of the throat, $\tau=0$ and discuss later the possibility of having the strings at other places. Consider $\theta= \theta_{1} =-\theta_{2}$ and $\phi=\phi_{1} = - \phi_{2}$. At $\psi=0$, this $(\theta, \phi)$ coordinate describes the shrinking $S^{2}$. 
This $(\theta, \phi)$ coordinate describes an $S^{2}$ inside the $S^{3}$ for $\psi=\pi$. We shall fix $\psi$ by minimizing the tension.

The metric Eq.(\ref{h(tau)}) at the tip is
\ba
\label{metric0}
ds^2 \sim  \h\, \eta_{\mu \nu} dx^{\mu} dx^{\nu}+
 b\, g_s M \alpha'(d\psi^2 +\sin^2 \psi\, d \oo_2 ^2)
\ea
where $h_A$ is the warp factor at the bottom of the throat 
and $b= 2^{2/3} 3^{-1/3} I(0)^{1/2}=0.93$ is a numerical constant in the KS solution.
Here $\psi$ is the usual azimuthal coordinate in a $S^3$ ranging from $0$ to $2\pi$, it is half of the coordinate $\psi_{T^{1,1}}$ defined in Eq.(\ref{6t11}). We also
absorbed $\alpha'$  in $x^{\mu}$ such that all coordinate in Eq.(\ref{metric0}) are dimensionless.

There are $M$ units of RR 3-forms $F_3$
on the non-vanishing $S^3$ cycle at the tip of the throat. Its associated two form is given by
\ba
\label{C2}
C_2=  M \alpha' \,  \left(\psi-\frac{\sin (2\psi)}{2} \right) \sin \theta\, d \theta \, d\phi \, .
\ea
At the bottom of the KS solution $B_{ab}=0$. 
The D3-brane is wrapped on an $S^2$ inside $S^3$ at the bottom of the throat.
Using metric (\ref{metric0}) and the background $C_2$ from Eq.(\ref{C2}) in our action (\ref{def}), we find
\begin{align}
\label{dd}
\dd & = 4 \pi \mu_3 g_s^{-1} \left [  (b\, g_s M \alpha')^2 \sin^4 \psi +(\frac{q\l}{2})^2  \right]^{1/2}\, ,\nonumber\\
& = \frac{1}{\l}\sqrt{\left(\frac{bM}{\pi}\right)^2\sin^4\psi +\frac{q^2}{g_s^2}}.
\end{align}
Also, 
\begin{align}
\label{oo}
\oo & = 4 \pi M \alpha' \mu_3 \lambda \left( \left(\psi -\frac{\sin(2\psi)}{2}\right) + \l C_0 \frac{q}{2}\right), \nonumber\\
& = qC_0 +\frac{M}{\pi}\left(\psi -\frac{\sin2\psi}{2}\right)\, .
\end{align}
Note that $C_0$ is zero to leading order in the KS background and so it is understood to be perturbatively small in our formula.
Using equations (\ref{dd}), (\ref{oo}) together with $D=4\pi\F^{10}=-4\pi\F^{01} = p$ in Hamiltonian 
Eq.(\ref{Ham}) we get 
\ba
{\cal{H}} = \frac{\h}{\l} \sqrt{ \frac{q^2}{g_s^2} + \frac{b^2M^2}{\pi^2} \sin^4 \psi+\left[ \frac{M}{\pi}
\left(\psi -\frac{\sin2\psi}{2}\right)  - (p-q C_0) \right]^2}\, .
\ea
In particular we note that the warp factor comes correctly as a overall pre-factor for the tension.
After minimization we find
\ba
\label{min1}
\frac{M}{\pi}\left(\psi +\frac{b^2-1}{2} \sin2\psi\right) - (p-qC_0) = 0 \, .
\ea
The tension of the $(p,q)$ bound state is
\ba
\label{pq}
{\cal{H}}_{min} =T_{(p,q)} = \frac{\h }{2\pi\ap}\sqrt{\frac{q^2}{g_s^2} + \frac{b^2 M^2}{\pi^2} \sin^2 \psi
\left(1+(b^2-1) \cos^2 \psi     \right)}
\ea
where $\psi$ is determined by the minimization condition (\ref{min1}). 

Note that $1-b=0.06734$ is quite small. For all practical purposes we may drop the $(b^{2}-1)$ term
in Eq.(\ref{min1}), so the calculations simplify and 
\ba
\label{pq2}
T_{(p,q)}  = {\cal{H}}_{min} \simeq  \frac{\h}{2 \pi \alpha'} \sqrt{\frac{q^2}{g_s^2}+
\left(\frac{b M}{\pi}\right)^2 \sin^2\left(\frac{\pi (p-qC_0)}{M}\right) }\, \, .
\ea
This expression is numerically within a percent of the more precise minimization result using Eq.(\ref{min1}). It also reproduces all the known limits.
If $p=0$, this gives the tension of $q$ D1-strings. If $q=0$, this gives the tension of $p$ F-strings as obtained by \cite{Herzog:2001fq}. On the other hand, if we consider the limit $M \rightarrow \infty$, then we obtain
\ba
\label{flatpq}
T= \frac{h_{A}^{2}}{2 \pi \alpha'} \sqrt{\frac{q^2}{g_s^2} + b^{2}(p-qC_0)^2}
\ea
Setting $b=1$ and $h_{A}=1$ yields the tension of $(p,q)$ strings in flat space-time, Eq.(\ref{flat}), where $C_{0}$ has also been set to zero.
 
\section{Discussion}

In Figure 1, the key paths are summarized in getting the $(p,q)$ string tension in a warped deformed conifold. It is clear that one can obtain the answer via a variety of possible paths. Note that for the D-string case the  $S^{2}$ cycle shrinks to zero size at the bottom of the throat, one can say that the D-string does not really blow up but the picture is nevertheless useful.

\begin{figure}[h]\label{complete1}
\centering
\psfrag{D3+ptF01}{D3+p$\F^{01}$}
\psfrag{D3+qF23}{D3+q$F_{23}$}
\psfrag{D3+qF23+ptF01}{D3+q$F_{23}$+p$\F^{01}$}
\psfrag{D3+qtF01}{D3+q$\F^{01}$}
\psfrag{D3+pF23}{D3+p$F_{23}$}
\psfrag{D3+pF23+qtF01}{D3+p$F_{23}$+q$\F^{01}$}
\psfrag{q D1}{qD1}
\psfrag{p D1}{pD1}
\psfrag{q F1}{qF1}
\psfrag{p F1}{pF1}
\psfrag{(p,q)}{(p,q)}
\psfrag{(q,p)}{(q,p)}
\psfrag{S}{S}
\psfrag{GHK}{GHK}
\psfrag{HK}{HK}
\psfrag{C2}{$C_2$}
\psfrag{? C2}{$C_2$}
\psfrag{? B2}{$B_2$}
\psfrag{B2}{$B_2$}
\includegraphics[width=10cm]{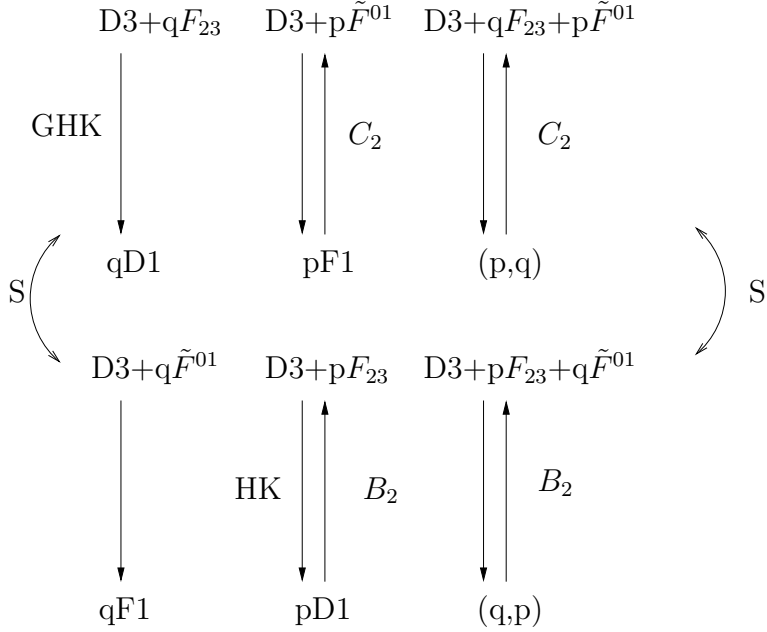}
\caption{The vertical arrows going down are just the result of dimensional reduction after integrating the 2-cycle over which the D3-brane wraps. Vertical arrows going up are the dielectric effect (blow-up) and we show in each case what is the background field creating the blow up. We then show the S-dual picture. GHK is the work of Gubser, Herzog and Klebanov Ref\cite{Gubser:2004qj} and HK refers to the work of Herzog and Klebanov Ref\cite{Herzog:2001fq}. }
\end{figure}
 
One can see in Figure 1 that an F-string is obtained from a D3-brane with electric flux which imply that, in the presence of RR fluxes, the F-string blows up in a D3-brane.  This is expected and it has been shown through matrix theory techniques that the F-string would indeed blow up in such a background \cite{Schiappa:2000dv, Brecher:2001dr} (see also \cite{Emparan:2001rp}).
Presumably the $(p,q)$ string obtained in this paper would blow up in a D3-brane with both magnetic flux and electric flux.  

It is an interesting question to see what the tension of $(p,q)$ strings would be in a general 
background. A reasonable guess motivated by Eq.(\ref{Ham}) is
\begin{align}
\label{Ham2}
{\cal{H}} & =  \sqrt{T_{D1}^2 + T_{F1}^2}
\end{align}
This expression works for the warped geometry and for the flat space-time. The latter just come from the SUSY algebra and is exact (because it is protected by supersymmetry) versus the former which comes from the form of the DBI action and has corrections. The corrections to the DBI action involves derivatives of the field strength over the string length.  We have shown that quite generally (assuming only $B_{01} =0$) we need constant fluxes on the D3-brane to leading order to get the correct charges. Therefore these corrections are expected to be very small.  This fact leads us to believe that the form (\ref{Ham2}) for the tension may hold quite generally with or without supersymmetry. 

Although the form of our formula Eq.(\ref{pq2}) might be very similar to the flat case Eq.(\ref{flatpq}), there are important differences coming in from the fluxes.  We have seen that a $C_2$ background comes into play in the F-string tension and analogously a $B_2$ background would affect the D-string tension.  The most notable effect from background of fluxes is to make the F-strings non-BPS and valued in $\mathbb{Z}_M$ for $M$ units of fluxes.  This has been shown to hold also for F-strings in 
 Maldacena-Nunez background \cite{Maldacena:2000yy, Herzog:2001fq}. Another important consequence of fluxes is to make the F- and D-string tensions incommensurate.  The $(p,q)$ binds even if $p$ and $q$ are not coprime. This is because the F-strings are $\mathbb{Z}_M$ charged.

It is expected that once one embeds the KS background in a compactified manifold the D-string should become valued in $\mathbb{Z}_K$.  One can see this from the gauge theory side. Once the KS solution is embedded in a compactification, the $U(1)$ baryon symmetry is gauge and it is broken down to $\mathbb{Z}_K$. The axionic string coming from this symmetry breaking should be $\mathbb{Z}_K$ charged. This is analogous to the picture in Ref.\cite{Leblond:2004uc} where it is shown how a global D1-string becomes local after compactification in the presence of a D3-brane. 
It is expected that, as local strings, the D-strings could break on monopoles \cite{Polchinski:2005bg}.  
On the gravity side, we have a dual B-cycle in the compactified theory with $K$ units of $H_3$ on it. This is exactly what we expect to get $\mathbb{Z}_K$ but because this cycle has a varying warp factor along it, all the D-strings are attracted to the bottom of the throat and they cannot get out. We speculate that subleading corrections to the warp factor might provide the key to get the $\mathbb{Z}_K$ expected from the gauge theory side.  Finally, the interpretation of the (p,q) string in the gauge theory dual can be seen as a bound state of confining strings and axionic strings but this need further study.

In the context of brane inflation, cosmic defects and especially cosmic superstrings appear quite naturally  \cite{Jones:2002cv, Sarangi:2002yt, Copeland:2003bj, Becker:2005pv}. 
In Type IIB theory, both the warp factor and the Dirac-Born-Infeld action enhance the number of e-folds of inflation needed to explain the flatness and the horizon problems. 
Heating after brane inflation is also quite feasible when the D3-anti-D3-brane annihilation takes place in one throat, while the standard model branes are somewhere else \cite{Firouzjahi:2005qs,Chen:2006ni}.
This annihilation releases the energy to heat up the universe to start the hot big bang epoch and to produce all defects that are permitted; in particular, F- and D-strings are expected to be copiously produced. Some of these superstrings will appear with cosmic sizes, as cosmic strings that are expected to evolve into a scaling cosmic string network. We note that $M$ F-strings will bind into a point; that is, they can bind into a ``baryon'' like point-like defect, with a mass $\sim M h_{A}/\sqrt{\ap}$. The cosmological evolution of such a cosmic string network will be interesting to study. This evolution might be quite non-trivial. Not only are the string breaking on baryons but their charge depends on where they are in the throat. At each step in the cascade the flux $F_3$ changes by one unit and so as a cosmic string moves randomly in the throat it could change from being $\mathbb{Z}_M$ charged to being $\mathbb{Z}_{M-1}$ charged for example.

A generic flux compactification will have a number of axions. One expects a new type of strings charged under each axion field. 
So it is likely that there are a variety of other strings besides the F- and the D-strings.
The resulting stringy bound states can be very rich. The $(p,q)$ string tension spectrum discussed in this note gives us a glimpse of what is possible.

\vspace{0.5cm}

\noindent {\bf \large {Acknowledgment}}

We thank Girma Hailu, Chris Herzog, Igor Klebanov and Sarah Shandera for valuable discussions. 
This work was supported by National Science Foundation under Grant No. PHY-009831.

\providecommand{\href}[2]{#2}\begingroup\raggedright\endgroup

\end{document}